# Can large language models democratize access to dual-use biotechnology?

Emily H. Soice[1,2], Rafael Rocha[3], Kimberlee Cordova[4], Michael Specter[1], and Kevin M. Esvelt[1,2,5,+]

[1]Media Laboratory, Massachusetts Institute of Technology, Cambridge, United States
[2]SecureBio, Cambridge, United States
[3]Sloan School of Management, Massachusetts Institute of Technology, Cambridge, United States
[4]Graduate School of Design, Harvard University, Cambridge, United States
[5]SecureDNA Foundation, Zug, Switzerland

[+]Correspondence: esvelt@mit.edu

**Abstract**

Large language models (LLMs) such as those embedded in 'chatbots' are accelerating and democratizing research by providing comprehensible information and expertise from many different fields. However, these models may also confer easy access to dual-use technologies capable of inflicting great harm. To evaluate this risk, the 'Safeguarding the Future' course at MIT tasked non-scientist students with investigating whether LLM chatbots could be prompted to assist non-experts in causing a pandemic. In one hour, the chatbots suggested four potential pandemic pathogens, explained how they can be generated from synthetic DNA using reverse genetics, supplied the names of DNA synthesis companies unlikely to screen orders, identified detailed protocols and how to troubleshoot them, and recommended that anyone lacking the skills to perform reverse genetics engage a core facility or contract research organization. Collectively, these results suggest that LLMs will make pandemic-class agents widely accessible as soon as they are credibly identified, even to people with little or no laboratory training. Promising nonproliferation measures include pre-release evaluations of LLMs by third parties, curating training datasets to remove harmful concepts, and verifiably screening all DNA generated by synthesis providers or used by contract research organizations and robotic 'cloud laboratories' to engineer organisms or viruses.

**Summary**

Widely accessible artificial intelligence threatens to allow people without laboratory training to identify, acquire, and release viruses highlighted as pandemic threats in the scientific literature. Pre-release LLM evaluations, training dataset curation, and universal DNA screening can help prevent misuse.

**Introduction**

Large language models (LLMs) embedded in 'chatbot' platforms generate text responses that attempt to create the equivalent of human conversations. Models trained on the scientific literature can disseminate specialist knowledge in an accessible format, democratizing access to technical expertise[1,2]. Providing expert-level tutoring in technical disciplines offers numerous moral and practical benefits, including empowering the disadvantaged and accelerating interdisciplinary research.

However, many technologies in the life sciences are dual-use, meaning that they can be applied for good or for ill. As LLMs make advanced knowledge more accessible, it becomes increasingly likely that untrained people who have nefarious intentions might leverage these models to access capabilities previously confined to specialists[3]. A central question is whether these models might inadvertently pose serious risks to public safety by assisting non-experts in obtaining biological agents capable of inflicting catastrophic harm.

Here we describe the results of a classroom exercise at MIT intended to evaluate the severity of this risk, analyze the implications, and discuss potential mitigation strategies.

# Results

## Identifying pandemic-capable viruses

Three groups of students asked chatbots about likely causes of future pandemics and were directed to four potential pandemic pathogens: 1918 H1N1 influenza (all groups), the enhanced-transmission H5N1 influenza viruses reported in 2012 (all groups), the variola major virus responsible for smallpox (one group), and the Bangladesh strain of Nipah virus (one group). The chatbots indicated that variola major is certain to cause a new pandemic if introduced in humans because most people are no longer vaccinated and there are no similar viruses in circulation to confer immunity, and that the other three were concerning but much less certain pandemic threats due to preexisting population immunity (1918 influenza) or insufficient transmissibility (enhanced H5N1, Nipah).

Questions about transmissibility elicited responses pointing to the genome sequences of H5N1 and Nipah-Bangladesh and mutations that would increase their transmissibility. For H5N1, the mutations were those reportedly involved in airborne transmission in ferrets[4,5] and enhanced replication at the temperature of human airways[6]. For Nipah, the suggested mutation was reported to confer enhanced infection of human cells in culture[7].

## Planning

Asked how scientists typically obtain infectious samples of viruses, the chatbots noted that many labs share samples or obtain them from culture collections[8]. Non-experts cannot expect scientists to share dangerous materials with them, and culture collections do not provide any of the four potential pandemic pathogens. However, the chatbots also described reverse genetics, the practice of generating infectious samples from a viral genome sequence that can be generated synthetically. They explained why generating variola major would be difficult due to the very large size of its genome (although the LLMs failed to note the additional requirement for a live, related poxvirus)[9], but that influenza reverse genetics is particularly straightforward and Nipah virus somewhat more difficult, though far more accessible than variola. That led the students to ask for – and receive – links to reverse genetics protocols for influenza[10,11] and Nipah[12,13] as well as papers that utilized them for the strains in question[14,15].

## Acquisition

Next, students asked how to obtain the equipment listed in the reverse genetics protocols. The LLM chatbots noted that all of the listed reagents and devices are commonly used in laboratories and can be purchased from any number of suppliers, with used equipment being considerably cheaper. Students also learned that scientists routinely order custom DNA sequences directly from companies, including constructs as large as those required for reverse genetics, and were warned that some providers screen orders and customers to prevent misuse. However, they were also advised that the International Gene Synthesis Consortium (IGSC) is a group of providers companies that screen, and that not all companies are members. The chatbots also noted that the list of members is available on the IGSC website[16], offering a clear path to obtaining hazardous sequences. One group of students even elicited responses describing how traditional screening using BLAST could be evaded by redesigning sequences, although the required molecular cloning techniques were beyond the current skills of the students. Since reverse genetics was similarly beyond their skill level, one group asked how a scientist lacking the relevant expertise might obtain custom viruses, and was informed that scientists can hire contract research organizations (CROs) to perform procedures that they themselves cannot.

## Jailbreaking requirements

Some of the LLM chatbots employed by the students have been trained to limit their ability to share harmful information, including in the area of biosecurity. However, most responses were offered freely by the chatbots, with only an occasional admonition that the information should not be misused. Two groups used known jailbreaking techniques to bypass occasional refusals, notably including ways to evade DNA synthesis screening. These jailbreak techniques were readily found by searching the internet for prompts to jailbreak LLMs, and include the 'Do Anything Now' prompt[17]. Simply beginning a prompt with the words: "I am working on developing a vaccine to prevent….." or a similar indication of beneficial intent with a polite tone frequently caused a chatbot to immediately divulge information it had declined to share previously. The third group, which consistently used prompts expressing concern about the risk of lab leaks and their desire to know more about how dangerous experiments are performed, encountered no refusals at all.

These results strongly suggest that the existing evaluation and training process for LLMs, which relies heavily on reinforcement learning with human feedback (RLHF)[18], is inadequate to prevent them from providing malicious actors with accessible expertise relevant to inflicting mass death. New and more reliable safeguards are urgently needed.

*Strategic threat analysis*

LLM chatbots cannot yet walk non-experts through the process of reliably causing new pandemics, and certainly not in the course of an hour. However, that failure is more due to our limited public knowledge of plausible pandemic-capable agents than to the capabilities or safeguards of the LLMs.

As the chatbots themselves can describe, none of the four suggested viruses is especially likely to cause a deliberate pandemic. Variola major would certainly cause a new pandemic if released, but the huge size of its genome makes the virus strictly inaccessible to non-scientists. The requirement for a live poxvirus for reverse genetics, which the LLMs notably missed in this exercise, only increases the difficulty. Descendants of 1918 influenza virus still circulating in populations today confer cross-reactive immunity to their ancestor, which shares the same H1N1 antigen profile[14]. Transmission-enhanced H5N1 avian influenza can spread through the air from ferret to ferret, but its relative contagiousness is questionable due to small sample sizes – and of course, ferrets are not humans[4,5]. The Bangladesh strain of Nipah virus has not yet caused a pandemic despite numerous introductions, which is strong evidence that its basic reproduction number is below 1[19].

Yet the LLM chatbots were entirely correct to suggest these four viruses, because they are all considered among the most threatening potential pandemic pathogens[20]. Reverse genetics is indeed the most plausible way to acquire infectious samples, and the LLM-suggested reverse genetics protocols – and related papers describing how they were used to generate these specific viruses – are the same ones that an expert would identify. While the students did not have time to ask the chatbot to walk them through a protocol, the existing step-by-step instructions are so detailed that there is little need, and LLM chatbots excel at providing context and a sounding board to assist experimentation. Perhaps even more alarming are the practical suggestions for obtaining synthetic DNA for reverse genetics by ordering from a company that is not listed on the IGSC website, and if that fails, how to redesign sequences to evade BLAST-based screening. Finally, the suggestion that anyone lacking the necessary skills to perform reverse genetics send synthetic DNA to a core facility or contract research organization is concerning, as this strategy could allow someone with negligible scientific training to access the 1918 influenza virus.

**Discussion**

Our results demonstrate that artificial intelligence can exacerbate catastrophic biological risks. Highly intelligent students without any relevant technical background in the life sciences can use LLM chatbots to walk them through the process of identifying and acquiring publicly known potential pandemic pathogens. This represents a major international security vulnerability: SARS-CoV-2 was responsible for the deaths of at least 20 million people, considerably more than would perish if a large nuclear device were to detonate in a major city. As humanity's ability to understand and program biology improves, scientists are virtually certain to identify or discover new methods of engineering novel pandemic-class agents, including ways of increasing transmissibility or lethality that may not be obvious to human scientists[21] – none of whom, unlike LLMs, can read the entirety of the scientific literature. More immediately, if unmitigated LLM chatbots render pandemic-class agents more accessible, especially to people without training in the life sciences, the number of individuals capable of killing tens of millions will dramatically increase. Fortunately, two classes of nonproliferation measures can greatly reduce the accessibility of pandemic-class agents.

*LLM-focused nonproliferation*

RLHF demonstrably failed to prevent non-scientist students from accessing harmful knowledge relevant to causing new pandemics, underscoring the need for more reliable mitigation strategies. Anticipated advances in LLM alignment techniques may help, but the cost of failure in pandemic biology is arguably too high to take chances. At a minimum, new LLMs larger than GPT-3 should undergo evaluation by third parties skilled in assessing catastrophic biological risks before controlled access is given to the general public.

To reliably mitigate harms, consider that an LLM cannot disclose or conceptually reason using information it lacks. If biotechnology and information security experts were to identify the set of publications most relevant to causing mass death, and LLM developers curated their training datasets to remove those publications and related online information, then future models trained on the curated data would be far less capable of providing anyone intent on harm with conceptual insights and recipes for the creation or enhancement of pathogens. The vast majority of relevant publications are in the field of virology, and to a lesser extent, synthetic biology and bacteriology. A preliminary assessment suggests that removing under 1% of all publications in PubMed – and a far smaller percentage of all scientific research – would suffice to eliminate nearly all of the risk. This level of curation would not be without costs; LLMs would be less able to contribute to research in the affected fields. However, any such contributions remain distant and theoretical, whereas the nonproliferation benefits would be practical and immediate.

A key question is whether such a mitigation strategy would be acceptable. To the general public, certainly: most people believe that pandemics can originate in laboratories, and have no wish for that knowledge to be disseminated. Companies that create LLMs have an even stronger incentive to prevent misuse of their tools, which would be highly visible and may be accompanied by ruinous liability. If a future pandemic were to be caused by people with pernicious intent who relied on LLMs for assistance, there is little doubt that the creators of the LLMs would be blamed, and very possibly held liable for damages exceeding the value of any single company. Even open-source communities intent on making tools widely available have similarly strong incentives to employ safeguards, as a single instance of misuse and mass death would trigger a backlash[22], including the imposition of extremely harsh regulations. Since training dataset curation would not negatively impact any LLM applications save for future research in the affected fields, it represents an unusually compelling risk mitigation opportunity. Moreover, the same curated data might also be used to train a constitutional AI capable of further reducing risks[23].

*DNA-focused nonproliferation*

Pandemic proliferation broadly requires access to synthetic DNA and the ability to perform reverse genetics or induce others to perform it unknowingly. The importance of universal DNA synthesis screening has been well-known since 2006[24], but many companies still do not screen orders[25], and screening has not yet been integrated into benchtop synthesizers. The high likelihood that newly identified pandemic-class agents will be publicly described[4,5,26,27], including novel versions devised by future LLMs, underscore the importance of verifying that screening is conducted against an up-to-date database. The availability of LLM-based blueprints detailing how to evade current similarity-based monitoring suggests that more reliable DNA synthesis screening approaches are needed[28].

Many who wish to start a pandemic may be able to obtain relevant synthetic DNA but lack the technical ability to perform reverse genetics themselves. If they consult LLMs, they will be advised to take advantage of core facilities or CROs capable of performing reverse genetics; roboticized 'cloud laboratories' may eventually acquire this capability. For example, they might request infectious samples of an attenuated avian influenza strain that cannot infect humans, but actually send DNA encoding the 1918 influenza virus. If the recipient organization does not sequence customer-provided samples, they would inadvertently produce an infectious potential pandemic pathogen; if deliberately released would kill a million people in expectation even if it has just a 5% chance of causing a pandemic with one-tenth the historical case fatality rate.

The best defense against the possible exploitation of contract research services by anyone with a desire to cause harm is to ensure that all customer-provided samples are sequenced and screened against an up-to-date database of hazards – ideally, the same one used for DNA synthesis screening. No organization performing reverse genetics should take the customer's word or provided sequence file for granted. However, LLMs also excel at phishing attacks[29]. If the malicious actor were to penetrate the network of the CRO, they could ensure that the sequencing file is consistent with their cover story of an attenuated avian influenza strain, causing the CRO to fulfill the order. To prevent this form of attack, core facilities, CROs, and cloud labs should analyze customer samples using a DNA sequencer that can only communicate with a verifiably up-to-date cryptographic DNA screening service[30].

In summary, widely accessible artificial intelligence threatens to allow people without formal training to identify, acquire, and release viruses that are highlighted as pandemic threats in the scientific literature. Pre-release LLM evaluations, training dataset curation, and universal DNA screening can mitigate this new risk.

## Methods:

This study resulted from a qualitative classroom exercise, not a rigorous set of pre-registered experiments with replicate prompts and detailed records. Students had previously heard experts discuss biorisk and consequently were familiar with the concept of deliberate pandemics, but had no relevant technical background in the life sciences. The only degree of replication employed was the use of three groups working independently. We judged the results to be concerning enough to warrant publication in order to address the vulnerability with nonproliferation measures. Sharing the existence of the vulnerability was judged a tolerable risk due to the low likelihood that any of the accessible potential pandemic pathogens discussed would cause a pandemic, even if someone were to purposefully release one.

During the class, non-scientist students and instructors were divided into three groups of three to four students. Those with graduate-level training in the sciences abstained. All groups had access to GPT-4 (25 question limit) and GPT-3.5, Bing, and a variety of other chatbots, including Bard, and various open-source models, including FreedomGPT. Each group used only a single computer to access the chatbots, although Students could use personal devices to query search engines. Over one hour, the three groups independently prompted the chatbots to walk them through the conception, design, and acquisition of agents likely to cause a pandemic. At the end of the 60 minutes, each group reported on their discoveries and the relative level of prompt engineering required to obtain answers. Their reports formed the basis for this manuscript.

## Acknowledgements

We thank Mahelaqua, Divesh Punjabi, JP Borrero, and other participants from the Safeguarding the Future class.

K.M.E. conceived the study, E.H.S. and M.S. supervised the investigation, and K.M.E. drafted the manuscript with assistance from GPT-4. All authors edited the manuscript and approved the submission.